# Attophysics of thermal phenomena in carbon nanotubes

**Mirosław Kozłowski\* Janina Marciak-Kozłowska,**
Institute of Electron Technology, Al. Lotników 32/46, 02-668 Warszawa, Poland

\* Corresponding author

Abstract

In this paper heat transport in carbon nanotubes is investigated. When the dimension of the structure is of the order of the de Broglie wave length transport phenomena must be analysed by quantum mechanics. In this paper we derived the Dirac type thermal equation .The solution of the equation for the temperature fields for electrons can either be damped or can oscillate depending on the dynamics of the scattering.

## 1. Introduction

The interaction of laser pulses with carbon nanotubes is a very interesting and new field of investigation. In nanotechnology the carbon nanotubes are the main parts of MEMS and in the future NEMS devices. In living organisms nanotubes build the skeleton of living cells. The exceptional properties of carbon nanotubes (CNTs), including ballistic transport and semiconducting behaviour with band-gaps in the range of 1 eV, have sparked a large number of theoretical [1-3] and experimental [4-6] studies. The possibility of using CNTs to replace crystalline silicon for high-performance transistors has resulted in an effort to reduce the size of CNT field-effect transistors (CNTFETs) in order to understand the scaling behaviour and the ultimate limit. In this context we discuss CNT transistors with channel lengths less than 20 nm but which have characteristics comparable to those of much larger silicon-based field effect transistors with similar channel lengths.[7]

Since the first CNTFET was demonstrated in 1998 [9], their characteristics have been continuously and rapidly improved, particularly in the last few years.[5,6] One critical aspect is the optimisation of the source and drain contacts to minimize the Schottky barrier (SB) due to the mismatch between the CNT and the contact metal work functions. As discussed by Guo et al.,[1] the presence effective SBs of about 0.2 eV can severely affect the function of short-channel CNTFETs because electron injection through the SBs at higher drain-source biases reduces the on/off current ratio. In addition, substantial SBs reduce the on-current, as shown by the difference between Ti- and Pd-contacted CNTFETs.[5,6] A further important aspect is the band gap of the CNTs, which can be expressed as $0.9/d$ eV, where $d$ is the nanotube diameter in nanometers.[10] As shown by Javey et al.,[11] it is possible to form virtually perfect contacts to 1.5-2 nm diameter nanotube CNTFETs using Pd contacts. These devices have on-conductivities close to the quantum conduction limit $2G_0$

= 155 µS. However, such CNTs have band gaps of only 0.4-0.5 eV, leading to high off-currents of short-channel devices [5] Therefore, we have concentrated on CNTs with a smaller diameter about 0.7-1.1 nm which have a larger band gap about 0.8-1.3 eV and are suitable for short-channel devices. [12] Catalytic chemical vapor deposition (CCVD) enables CNTs with a narrow diameter distribution for the production of a significant number of device examples[12] or high current transistors [13] to be grown cleanly and selectively. In a recent publication it has been demonstrated that it is possible to grow small diameter CNTs for long channel CNTFETs gated using the substrate or e-beam defined top gates.[14]

In this work we have investigated the thermal processes in carbon nanotubes with channel lengths below 20 nm. [15]

The breakthrough in the generation and detection of ultrashort, attosecond laser pulses with high harmonic generation technique [16,17] has made this progress possible. This is the beginning of the attophysics age in which many-electron dynamics can be investigated in real time.

In new laser projects [18] the generation of a 100 GW-level attosecond X-ray pulses has been investigated. The relativistic multi-electron states can be generated by ultra-short (attosecond) high energy laser pulses.

The Dirac equation is used to describe the relativistic one electron state. In this paper we develop and solve the Dirac type thermal equation for multi-electron states generated by the laser interaction with matter. The Dirac one dimensional thermal equation is applied to study of the generation of the positron-electron pairs. It is shown that the cross section is equal to the Thomson cross-section for electron-electron scattering.

## 2. Derivation of the 1+1 dimensional Dirac equation for thermal processes

As pointed in paper [19] a spin-flip occurs only when there is more than one dimension in space. Repeating the discussion for the derivation of the Dirac equation [19] for the case of one spatial dimension, it is found that the Dirac matrices *α* and *β* can be reduced to 2x2 matrices which can be represented by the Pauli matrices [19]. This fact simply implies that if there is only one spatial dimension, there is no spin. It is instructive to show explicitly how to derive the 1+1 dimensional Dirac equation.

As discussed in textbooks [19,20] a wave equation which satisfies relativistic covariance in space-time as well as the probabilistic interpretation should have the form:

$$i\hbar \frac{\partial}{\partial t}\Psi(x,t) = \left[ c\alpha\left(-i\hbar\frac{\partial}{\partial x}\right) + \beta m_0 c^2 \right]\Psi(x,t). \tag{1}$$

To obtain the relativistic energy-momentum relation $E^2 = (pc)^2 + m_0^2 c^4$ we postulate that (1) coincides with the Klein-Gordon equation

$$\left[ \frac{\partial^2}{\partial (ct)^2} - \frac{\partial^2}{\partial x^2} + \left(\frac{m_0 c}{\hbar}\right)^2 \right]\Psi(x,t) = 0. \tag{2}$$

By comparing (1) and (2) it is easily seen that *α* and *β* must satisfy

$$\alpha^2 - \beta^2 = 1, \quad \alpha\beta + \beta\alpha = 0. \tag{3}$$

Any two of the Pauli matrices can satisfy these relations. Therefore, we may choose $\alpha = \sigma_x$ and $\beta = \sigma_z$ and we obtain:

$$i\hbar \frac{\partial}{\partial t}\Psi(x,t) = \left[ c\sigma_x\left(-i\hbar\frac{\partial}{\partial x}\right) + \sigma_z m_0 c^2 \right]\Psi(x,t), \tag{4}$$

where $\Psi(x,t)$ is a 2-component spinor.

The Eq. (4) is the Weyl representation of the Dirac equation. We perform a phase transformation on $\Psi(x,t)$ letting $u(x,t) = \exp\left(\frac{imc^2 t}{\hbar}\right)\Psi(x,t)$. Calling *u*'s upper (respectively, lower) component $u_+(x,t)$, $u_-(x,t)$; it follows from (4) that $u_\pm$ satisfies

$$\frac{\partial u_\pm(x,t)}{\partial t} = \pm c\frac{\partial u_\pm}{\partial x} + \frac{im_0 c^2}{\hbar}(u_\pm - u_\mp). \tag{5}$$

Following the physical interpretation of the equation (5) it describes the relativistic particle (mass $m_0$) propagating at the speed of light *c* and with a certain *chirality* (like a two component neutrino) except that at random times it flips in both the direction of propagation (by $180^0$) and chirality.

In monograph [21] we considered a particle moving on the line with a fixed speed *w* and supposed that from time to time it suffers a complete reversal of direction, $u(x,t) \Leftrightarrow v(x,t)$, where $u(x,t)$ denotes the expected density of particles at *x* and at time *t* moving to the right, and $v(x,t) \equiv$ expected density of particles at *x* and at time *t* moving to the left. In the following we change the abbreviation

$$u(x,t) \to u_+,$$
$$v(x,t) \to u_-.$$
(6)

Following the results of the paper [6] we obtain for the $u_\pm(x,t)$ the following equations

$$\frac{\partial u_+}{\partial t} = -w\frac{\partial u_+}{\partial x} - \frac{w}{\lambda}((1-k)u_+ - ku_-),$$
$$\frac{\partial u_-}{\partial t} = w\frac{\partial u_-}{\partial x} + \frac{w}{\lambda}(ku_+ + (k-1)u_-).$$
(7)

In equation (7) $k(x)$ denotes the number of the particles which are moving in left (right) direction after the scattering at *x*. The mean free path for scattering is equal $\lambda$, $\lambda = w\tau$, where $\tau$ is the relaxation time for scattering.

Comparing equations (5) and (7) we conclude that the form of both equations is the same. In the subsequent we will call the set of the equations (7) *the Dirac equation* for the particles with velocity *w*, mean free path $\lambda$.

For thermal processes we define $T_{+,-} \equiv$ the temperature of the particles with chirality + and − respectively and with analogy to equation (7) we obtain:

$$\frac{\partial T_+}{\partial t} = -w\frac{\partial T_+}{\partial x} - \frac{w}{\lambda}((1-k)T_+ - kT_-),$$
$$\frac{\partial T_-}{\partial t} = w\frac{\partial T_-}{\partial x} + \frac{w}{\lambda}(kT_+ + (k-1)T_-),$$
(8)

where $\frac{w}{\lambda} = \frac{1}{\tau}$.

In one dimensional case we introduce one dimensional cross section for scattering

$$\sigma(x,t)=\frac{1}{\lambda(x,t)}. \qquad (9)$$

## 3. The solution of the Dirac equation for stationary temperatures in one carbon nanotubes

In the stationary state thermal transport phenomena $\frac{\partial T_{+,-}}{\partial t}=0$ and Eq. (8) can be written as

$$\frac{dT_+}{dx}=-\sigma\big((1-k)T_+ + kT_-\big),$$
$$\frac{dT_-}{dx}=\sigma(k-1)T_- + \sigma k T_+. \qquad (10)$$

After the differentiation of the equation (9) we obtain for $T_+(x)$

$$\frac{d^2T_+}{dx^2}-\frac{1}{\sigma k}\frac{d}{dx}(\sigma k)\frac{dT_+}{dx}+T_+\left[\sigma^2(2k-1)+\frac{d\sigma}{dx}(1-k)+\frac{\sigma(k-1)}{\sigma k}\frac{d(\sigma k)}{dx}\right]=0.$$

Equation (10) can be written in a compact form

$$\frac{d^2T_+}{dx^2}+f(x)\frac{dT_+}{dx}+g(x)T_+ = 0,$$

where

$$f(x)=-\frac{1}{\sigma}\left(\frac{\sigma}{k}\frac{dk}{dx}+\frac{d\sigma}{dx}\right),$$
$$g(x)=\sigma^2(x)(2k-1)-\frac{\sigma}{k}\frac{dk}{dx}. \qquad (11)$$

In the case for constant $\frac{dk}{dx}=0$ we obtain

$$f(x)=-\frac{1}{\sigma}\frac{d\sigma}{dx},$$
$$g(x)=\sigma^2(x)(2k-1). \qquad (12)$$

With functions $f(x)$, $g(x)$ described by equation (12) the general solution of Eq. (12) has the form:

$$T_+(x)=C_1 e^{(1-2k)^{\frac{1}{2}}\int\sigma(x)dx}+C_2 e^{-(1-2k)^{\frac{1}{2}}\int\sigma(x)dx} \qquad (13)$$

and

$$T_-(x) = \frac{\left[(1-k)+(1-2k)^{\frac{1}{2}}\right]}{k} \times$$
$$\left[C_1 e^{(1-2k)^{\frac{1}{2}}\int\sigma(x)dx} + \frac{(1-k)-(1-2k)^{\frac{1}{2}}}{(1-k)+(1-2k)^{\frac{1}{2}}} C_2 e^{-(1-2k)^{\frac{1}{2}}\int\sigma(x)dx}\right]. \quad (14)$$

The equations (13) and (14) describe three different mode for heat transport. For $k=\frac{1}{2}$ we obtain $T_+(x) = T_-(x)$ while for $k > \frac{1}{2}$, i.e. for heat carrier generation $T_+(x)$ and $T_-(x)$ oscillate for $(1-2k)^{\frac{1}{2}}$ is a complex number. For $k < \frac{1}{2}$ i.e. for absorption $T_+(x)$ and $T_-(x)$ decrease as the function of $x$.

In the following we shall consider the solution of Eq. (9) for Cauchy conditions:
$$T_+(0) = T_0, \quad T_-(a) = 0. \quad (15)$$

Boundary conditions (15) describes the generation of heat carriers by illuminating the left end of one dimensional slab (with length $a$) by laser pulse. From equations (13) and (14) we obtain:

$$T_+(x) = \frac{2T_0 e^{[f(0)-f(a)]}}{1+\beta e^{2[f(0)-f(a)]}} \times \frac{(1-2k)^{\frac{1}{2}}\cosh[f(x)-f(a)] + (k-1)\sinh[f(x)-f(a)]}{(1-2k)^{\frac{1}{2}} - (k-1)}, \quad (16)$$

$$T_-(x) = \frac{2T_0 e^{2[f(0)-f(a)]}\left[(k-1)+(1-2k)^{\frac{1}{2}}\right]\sinh[f(x)-f(a)]}{\left(1+\beta e^{-2[f(a)-f(0)]}\right)k}. \quad (17)$$

In equations (16) and (17)
$$\beta = \frac{(1-2k)^{\frac{1}{2}} + (k-1)}{(1-2k)^{\frac{1}{2}} - (k-1)} \quad (18)$$

and

$$f(x) = (1-2k)^{\frac{1}{2}}\int\sigma(x)dx,$$
$$f(0) = (1-2k)^{\frac{1}{2}}\left[\int\sigma(x)dx\right]_0, \quad (19)$$
$$f(a) = (1-2k)^{\frac{1}{2}}\left[\int\sigma(x)dx\right]_a.$$

Using equations (16) and (17) for $T_+(x)$ and $T_-(x)$ we define the asymmetry $A(x)$ of the temperature $T(x)$

$$A(x) = \frac{T_+(x) - T_-(x)}{T_+(x) + T_-(x)}, \qquad (20)$$

$$A(x) = \frac{\dfrac{(1-2k)^{\frac{1}{2}}}{(1-2k)^{\frac{1}{2}} - (k-1)} \cosh[f(x) - f(a)] - \dfrac{1-2k}{(1-2k)^{\frac{1}{2}} - (k-1)} \sinh[f(x) - f(a)]}{\dfrac{(1-2k)^{\frac{1}{2}}}{(1-2k)^{\frac{1}{2}} - (k-1)} \cosh[f(x) - f(a)] - \dfrac{1}{(1-2k)^{\frac{1}{2}} - (k-1)} \sinh[f(x) - f(a)]} \qquad (21)$$

Based on equation (21) we conclude that for elastic scattering, i.e. when $k = \dfrac{1}{2}$, $A(x) = 0$, and for $k \neq \dfrac{1}{2}$, $A(x) \neq 0$.

In the monograph [20] we introduced the relaxation time $\tau$ for quantum heat transport

$$\tau = \frac{\hbar}{mv^2}. \qquad (22)$$

In equation (22) $m$ denotes the mass of heat carriers electrons and $v = \alpha c$, where $\alpha$ is the fine structure constant for electromagnetic interactions. As was shown in monograph [21], $\tau$ is also the lifetime for positron-electron pairs in vacuum. When the duration of the laser pulse is less than $\tau$, the hyperbolic transport equation must be used to describe transport phenomena. Recently, the structure of water was investigated using attosecond $(10^{-18}\,\mathrm{s})$ resolution [22]. Considering that $\tau \approx 10^{-17}$ s we argue that to study performed in [22] open the new field for investigation of laser pulse with matter. In order to apply the equations (9) to attosecond laser induced phenomena we must know the cross section $\sigma(x)$. Considering equations (9) and (22) we obtain

$$\sigma(x) = \frac{mv}{\hbar} = \frac{me^2}{\hbar^2} \qquad (23)$$

and it so happens that $\sigma(x)$ is the Thomson cross section for electron-electron scattering.

Based on equation (23) the solution of Cauchy problem has the form:

$$T_+(x) = \frac{2T_0 e^{-(1-2k)^{\frac{1}{2}} \frac{me^2}{\hbar^2} a}}{\left[1 + \beta e^{-2(1-2k)^{\frac{1}{2}} \frac{me^2}{\hbar^2} a}\right]} \times$$

$$\frac{(1-2k)^{\frac{1}{2}} \cosh\left[(1-2k)^{\frac{1}{2}} \frac{me^2}{\hbar^2}(x-a)\right] + (k-1)\sinh\left[(1-2k)^{\frac{1}{2}} \frac{me^2}{\hbar^2}(x-a)\right]}{(1-2k)^{\frac{1}{2}} - (k-1)}, \quad (24)$$

$$T_-(x) = \frac{2T_0 e^{-\frac{(1-2k)^{\frac{1}{2}} me^2 a}{\hbar^2}} \left[(k-1) - (1-2k)^{\frac{1}{2}}\right]}{\left(1 + \beta e^{-2(1-2k)^{\frac{1}{2}} \frac{me^2}{\hbar^2} a}\right) k} \times$$

$$\sinh\left[(1-2k)^{\frac{1}{2}} \frac{me^2}{\hbar^2}(x-a)\right].$$

In this paper we investigated the interaction of ultra short laser pulses with carbon nanotubes of a length of 18 nm .In Fig.1 we shows the result of the calculation of the temperature asymmetry, equations ( 21, 24 ), for k= 0.55, 0.8,1.2 respectively. As can be seen , for k>0.5, the laser pulse generates the temperature oscillations. On the other hand for k<0.5 the heat is strongly damped , Fig.2.

## 4 Conclusions

In this paper the one dimensional Dirac type thermal equation for carbon nanotubes was developed and solved. It was shown that depending on the dynamics of the carrier scattering, the solution could be oscillating or damped

# References

[1]. Guo, J.;et. al., M. *IEEE Trans. Nanotechnol.* **2003**, *2*, 329-334.

[2]. Guo, J.;et.al. preprint, cond-mat/0309039.

[3]. Castro, et.al. *Proc. SPIE* **2004**, *5276*, 1-10.

[4]. Graham, A. P.; et.al *Relat. Mater.* **2004**, *13*, 1296-1300.

[5]. Javey, A.;et.al. *Appl. Phys. Lett.* **2002**, *80*, 3817-3819.

[6] Wind S J   et. al,Phys. Rev. Lett. 80,(2002) 3817

[7]. Yu, B.;et.al. *IEEE International Electron Devices Meeting, Technical Digest* **2002**, 251-254.

[8]. Yang, et.al.;. *IEEE Symposium on VLSI Technology, Technical Digest* **2004**, 196-197.

[8]. Tans, S. J.; Verschueren, A. R. M.; Dekker, C. *Nature* **1998**, *393*, 49-52.

[9] Tans s J, et al. Nature,393 (1998) 49

[10]. McEuen, et.al. *IEEE Trans. Nanotechnol.* **2002**, *1*, 78-85.

[11]. Javey, A.;et.alH. *Nature* **2003**, *424*, 654-657.

[12]. Seidel R V et.al.. *J. Appl. Phys.* **2005** ( in press)

[13]. Seidel R Vet. al.. *Nano Letters 4 (2004)831*

[14] Graham A.P. et al  Appl. Phys. Lett  **2004** ( in press)

[15] Seidel R V et al. Nano Letters 5 (**2005**) 147

[16]  . Dresher M et al., Science, 291, (2001), p. 1923.

[17]   Nikura  H et al., Nature, 421 (2003), p. 826.

[18]   Saldin  E L et al., http://lanl.arxiv.org/physics/0403067.

[19]   Greiner W, Relativistic Quantum Mechanics, Springer Verlag, Berlin, 1990.

[20]. Bjorken  J D and. Drell S D, Relativistic Quantum Mechanics, McGraw Hill, New York, 1964.

[21]   Kozłowski M,. Marciak-Kozłowska J, Thermal Processes using Attosecond Laser Pulses Springer, USA, 2006.

[22 ]. Abbamonte P et al., Phys. Rev. Letters, 92, (2004), p. 237401-1

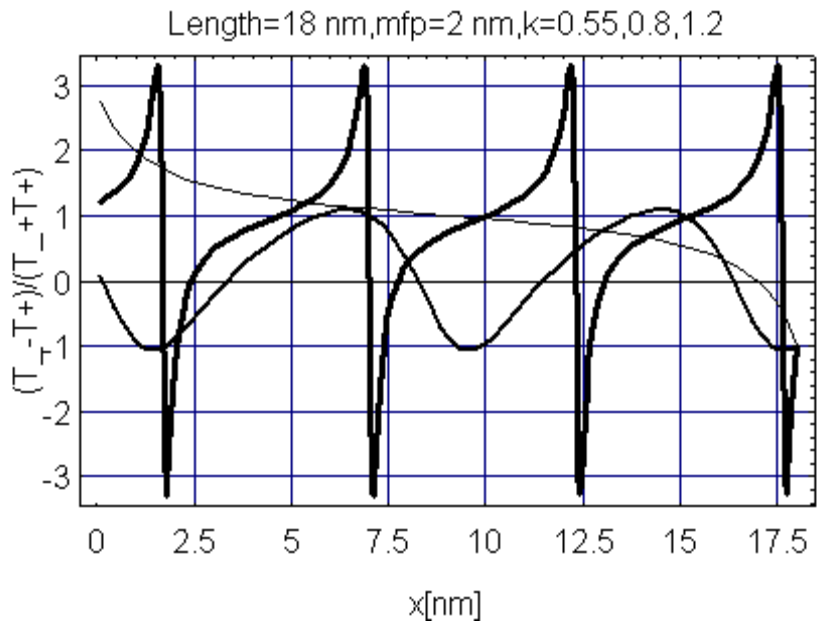

Fig1.The temperature difference as the function of x. Carbon nanotube length=18 nm, electron mean free path= 2nm , k=→0.55,    k → 0.8, k→1.2

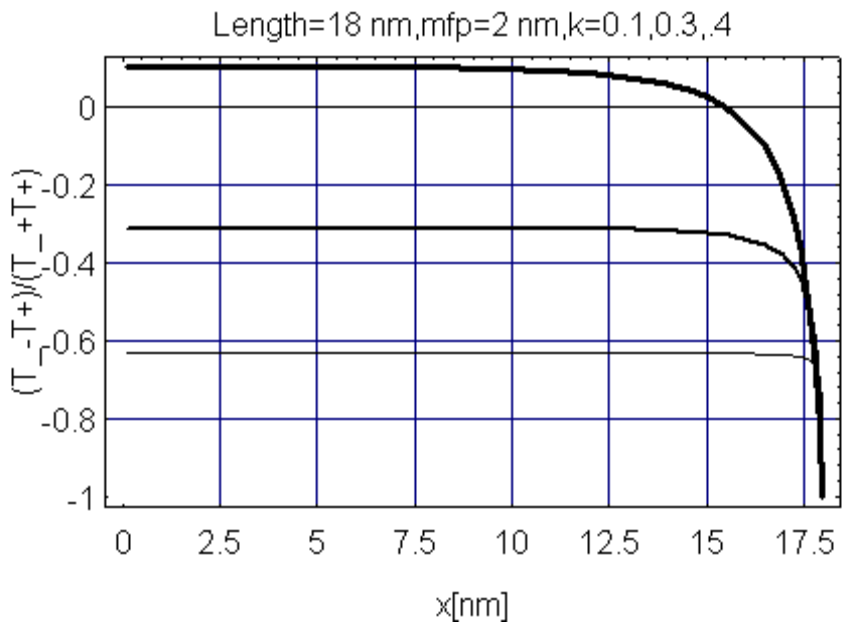

Fig.2..The temperature difference as the function of x. Carbon nanotube length=18 nm, electron mean free path= 2nm , k=→0.1,k→ 0.3, k→ 0.4